\documentclass[10pt]{iopart}
\usepackage{iopams,natbib,graphicx}

\def\beq{\begin{equation}}
\def\eeq{\end{equation}}
\renewcommand{\cite}[1]{\citep{#1}}
\newcommand{\la}{\lesssim}

\newcommand{\apj}{{\it Astrophys. J.}}

\newcommand{\apjl}{{\it Astrophys. J. Lett.}}
\newcommand{\mnras}{{\it Mon. Not. R. Astron. Soc.}}
\newcommand{\prl}{{\it Phys. Rev. Lett.}}

\newcommand{\rem}[1]{ }

\bibpunct[,]{(}{)}{;}{a}{}{,}
\begin{document}

\title{Magnetic fields and cosmic rays in GRBs.  A self-similar collisionless foreshock}

\author{Mikhail V. Medvedev$^1$\footnote{Also: Institute for Nuclear Fusion, RRC ``Kurchatov
Institute'', Moscow 123182, Russia}
and Olga V. Zakutnyaya$^2$}
\address{$^1$Department of Physics and Astronomy, University of Kansas, Lawrence, KS 66045}
\address{$^2$Space Research Institute (IKI), Moscow 117997, Russia}
\ead{medvedev@ku.edu}

\begin{abstract}
Cosmic rays accelerated by a shock form a streaming distribution of outgoing particles in the foreshock region. If the ambient fields are negligible compared to the shock and cosmic ray energetics, a stronger magnetic field can be generated in the shock upstream via the streaming (Weibel-type) instability. Here we develop a self-similar model of the foreshock region and calculate its structure, e.g., the magnetic field strength, its coherence scale, etc., as a function of the distance from the shock. Our model indicates that the entire foreshock region of thickness $\sim R/(2\Gamma_{\rm sh}^2)$, being comparable to the shock radius in the late afterglow phase when $\Gamma_{\rm sh}\sim1$, can be populated with large-scale and rather strong magnetic fields (of sub-gauss strengths with the coherence length of order $10^{17}\ {\rm cm}$) compared to the typical interstellar medium magnetic fields. The presence of such fields in the foreshock region is important for high efficiency of Fermi acceleration at the shock. Radiation from accelerated electrons in the foreshock fields can constitute a separate emission region radiating in the UV/optical through radio band, depending on time and shock parameters. We also speculate that these fields being eventually transported into the shock downstream can greatly increase radiative efficiency of a gamma-ray burst afterglow shock. 
\end{abstract}

\submitto{Astrophys. J.}

\section{Introduction}

Do gamma-ray bursts (GRBs) accelerate cosmic rays (CRs)? There are arguments in favor of such an idea \citep{DA04}. It is supposed that CRs are accelerated via the Fermi mechanism in which a particle crosses the shock many times and gradually gains its energy. For an ultra-relativistic shock, a steady-state universal power-law energy distribution of particles shall form \citep{Kirk+00,Acht+01}. The shock shall accelerate all particle species, but the electrons being much lighter than protons and ions will also loose their energy via synchrotron cooling. This radiation is thought to be observed as the delayed afterglow emission of GRB sources. A problem immediately arises here from a simple estimate: if the X-ray afterglow observed on a day timescale after the prompt burst is indeed the synchrotron radiation from the shock-accelerated electrons, then the pre-shock medium has to be highly magnetized with the fields of milligauss strengths \citep{LW06}. Thus, either the magnetic field is somehow generated in the shock upstream, or the conventional paradigm of the GRB afterglow needs revision.

In this paper, we present a self-similar model of the large region in front of a relativistic shock -- the foreshock. This region is populated with the shock-accelerated particles, which stream away from the shock into the collisionless ambient medium and generate magnetic field via a streaming (Weibel-type) instability. The model predicts the generation of strong, sub-gauss, magnetic fields in the entire foreshock whose thickness is $\sim R/(2\Gamma_{\rm sh}^2)$ and is comparable to the shock radius, $\sim10^{17}-10^{18}\ {\rm cm}$, in the afterglow phase a day or more after the explosion, when the shock is weakly relativistic or non-relativistic. The fields are sustained against dissipation by the anisotropy of newly accelerated particles. Moreover, these fields are relatively large-scale, with the coherence length being as large as a fraction of the foreshock size, $\sim10^{16}\ {\rm cm}$, which makes them effectively decoupled from dissipation. We speculate, that these mesoscale magnetic fields being ultimately advected into the shock downstream can significantly increase the radiative efficiency of GRB afterglows and, perhaps, explain the origin of the magnetic field in an external shock of a GRB. We remark here, however, that our study is analytical and cannot account for a number of nonlinear feedback effects of the generated fields and pre-conditioned external medium onto the shock structure and particle acceleration. Kinetic and hybrid computer modeling is essential for better and more accurate understanding of the foreshock structure.

\section{The model}

Overall, our model is as follows. A shock is a source of CRs which move away from it, thus forming a stream of particles through the ambient medium, say, the interstellar medium (ISM). If the ISM magnetic fields are negligible, i.e., their energy density is small compared to that of CRs, the streaming instability (either the pure magnetostatic Weibel or the mixed-mode electromagnetic oblique Weibel-type instability, depending on conditions) is excited and stronger magnetic fields are quickly generated. These fields further isotropize (thermalize) the CR stream. Since less energetic particles, having a greater number density and carrying more energy overall, are thermalized closer to the shock, the generated B-field will be stronger closer to the shock and fall off away from it, whereas its correlation length will increase with the increasing distance from the shock.  More energetic particles keep streaming because of their larger Larmor radii and produce the magnetic field further away from the shock. This process stops at distances where either the CR flux starts to decrease (because of the finite distance the CR particles can get away from a relativistic shock or because of the shock curvature causing CR density to decrease as $\propto r^{-2}$ if the shock is sub- or non-relativistic) or where the generated magnetic fields become comparable to the ISM field and the instability ceases. Thus, a large upstream region --- the foreshock --- is populated with magnetic fields. We now derive its self-similar structure. We work in the shock co-moving frame unless stated otherwise. 

Let's consider a relativistic shock moving along $x$-direction with the bulk Lorentz factor $\Gamma_{\rm sh}$; the shock is plane-parallel and lies in the $yz$-plane, and $x=0$ denotes the shock position. The shock continuously accelerates cosmic rays, which then propagate away from it into the upstream region. We conventionally assume that the CR distribution over the particle Lorentz factor is described by a power-law: 
\beq
n_{\rm CR}=n_0(\gamma/\gamma_0)^{-s}
\label{nCR}
\eeq
for $\gamma>\gamma_0$ and zero otherwise. Here the index $s=p-1$ is approximately equal to 1.2 for ultrarelativistic shocks and $n_0$ is the normalization.\footnote{Conventionally the distribution is given as $dn/d\gamma\propto\gamma^{-p}$ with $p$ being $\sim2.2-2.3$ for relativistic shocks; hence the density of particles of energy $\sim\gamma$ is $n(\gamma)\propto\gamma^{-p}\delta\gamma\propto\gamma^{-p+1}$.}
 We assume that the above energy distribution is the same everywhere in the upstream, that is, we neglect the nonlinear feedback of magnetic fields onto the particle distribution. The CR momentum distribution exhibits strong anisotropy: the parallel ($x$) components of CR momenta are much greater than their thermal spread in the perpendicular ($yz$) plane. Indeed, for a particle to move away from the shock, it should have the $x$-component of the velocity exceeding the shock velocity. Since both the shock and the particle move nearly at the speed of light, this puts a constraint on their relative angle of propagation to be less then $1/\Gamma_{\rm sh}$ in the lab (observer) frame. Hence, the transverse spread of the CR particle's momenta is $p_\perp\la p_\|/ \Gamma_{\rm sh} \ll p_\|$. This is also seen in numerical simulations \citep{Spit08}.

The CR particles propagate through the self-generated foreshock fields and scatter off them. Lower energy particles are deflected in the fields more strongly and, therefore, izotropize faster than the higher energy ones, as having larger Larmor radii.  At a position $x>0$ the CR distribution can roughly be divided into isotropic (themalized) component with $\gamma<\gamma_r(x)$ and streaming component with $\gamma>\gamma_r(x)$, where $\gamma_r(x)$ is the minimum Lorentz factor of the streaming particles at a location $x$; it is also the maximum Lorentz factor of the randomized component at this location. The streaming component is Weibel-unstable with a very short $e$-folding time $\tau\sim\omega_{p,{\rm rel}}^{-1}$, where $\omega_{p,{\rm rel}}=\left(4\pi e^2 \tilde n(\gamma)/m_p \gamma\right)^{1/2}$ is the relativistic plasma frequency, $\tilde n(\gamma)$ is the density of streaming particles of the Lorentz factor $\gamma$ (tilde denotes streaming particles). Note that the Weibel instability growth rate depends on $n$ of the lower density component -- cosmic rays, in our case -- measured in the center of mass frame of the streaming plasmas. For the lower-energy part of the CR distribution, the center of mass frame is approximately the shock co-moving frame, hence we evaluate the instability on the shock frame. This approximation is less accurate for the high-energy CR tail; however, the growth rate and the scale length are weak functions of the the shock Lorentz factor ($\propto\Gamma_{\rm sh}^{\mp 1/2}$), so the result will be accurate within an order of magnitude for all reasonable values of $\Gamma_{\rm sh}$ for GRB afterglows.  Here we use the proton plasma frequency because the CR electron Lorentz factors are about $m_p/m_e$ times larger, so they behave almost like protons \citep{Spit08}. The instability is very fast: it rapidly saturates (the fields cease to grow) in a few tens of $e$-folding times $\tau$, that is in few tens of inertial  lengths (also referred to as the ion skin length)  $c/\omega_{p,{\rm rel}}$ in front of the shock.  Thereafter the particles keep streaming in current filaments and the field around them amounts to $\xi_B\sim0.01-0.001$ or so of the kinetic energy of this group of particles: 
\beq
B^2(\gamma)/8\pi\sim\xi_Bm_pc^2\gamma \tilde n(\gamma),
\label{B-gamma}
\eeq 
where $\xi_B$ is the efficiency factor obtained from particle-in-cell (PIC) simulations; $\xi_B$ has the same meaning as the conventional $\epsilon_B$ parameter reserved here for the ratio of the total magnetic energy to the total kinetic energy of the shock and which, as is seen in PIC simulations, is larger than $\xi_B$ near the shock because of the nonlinear evolution and filament mergers. The correlation length of the field is of the order of the ion inertial length 
\beq
\lambda(\gamma)\sim c/\omega_{p,{\rm rel}}=\left(m_pc^2 \gamma/4\pi e^2 \tilde n(\gamma)\right)^{1/2}.
\label{lambda-gamma}
\eeq

These random fields deflect CR particles and ultimately lead to their isotropization. The deflection angle of the particle on a field coherence length scale in the self-generated field is
\beq
\theta\sim\delta p_\perp/p\sim(\lambda/c)\omega_B,
\label{theta-gamma}
\eeq
where $\omega_B=eB/\gamma m_p c$, $p$ is the particle momentum and $\delta p_\perp$ is it's transverse change. Using Eqs. (\ref{B-gamma}), (\ref{lambda-gamma}), we obtain:
\beq
\theta\sim eB(\gamma)\lambda(\gamma)/(\gamma m_pc^2)\sim\sqrt{2\xi_B}.
\eeq
Note that the deflection angle is independent of the particle's energy, as long as the field is produced by the particles of the same energy $\gamma$. The particles diffuse in the field and their rms deflection angle after transiting through a distance $x$ is $\Theta\sim\theta\sqrt{x/\lambda}$. The group of particles thermalizes when $\Theta\sim1$, i.e., at the distance from the shock:
\beq
x_r\sim\lambda/\theta^2\sim\lambda(\gamma)/(2\xi_B).
\label{x-gamma}
\eeq 
At this point, $x=x_r$, one has $\gamma=\gamma_r$ by definition; no field of the strength $B(\gamma_r)$ and the scale $\lambda(\gamma_r)$ can be produced at $x>x_r$. Similarly, one can estimate the randomization of the higher energy particles with $\gamma\gg\gamma_r$: $\theta(\gamma)\sim\sqrt{2\xi_B}(\gamma_r/\gamma)\ll\theta(\gamma_r)$, which means that these particles keep streaming through much larger distances $x\gg x_r$ and will produce the magnetic field further away from the shock. This field will be weaker and larger scale because of the lower density of the streaming particles $\tilde n(\gamma)\ll\tilde n(\gamma_r)$, according to Eqs. (\ref{nCR})--(\ref{lambda-gamma}). 

Finally, the number density of streaming CR particles at $\gamma_r$ is $\tilde n(\gamma_r)=n_0(\gamma_r/\gamma_0)^{-s}$. Therefore,
\beq
\lambda(\gamma_r)
\sim\left(m_pc^2 \gamma_0/4\pi e^2 n_0\right)^{1/2}(\gamma_r/\gamma_0)^{(1+s)/2}
\equiv\lambda_0(\gamma_r/\gamma_0)^{(1+s)/2},
\eeq 
where $\lambda_0$ is the inertial length of the lowest energy CR ``plasma". Inverting this expression yields:
\beq
\gamma_r\sim\gamma_0[\lambda(\gamma_r)/\lambda_0]^{2/(1+s)}
\sim\gamma_0(2\xi_Bx_r/\lambda_0)^{2/(1+s)}.
\label{gamma_r}
\eeq
Hereafter, the subscript ``$r$'' can be omitted without loss of clarity. 

In a steady state, this field is continuously advected toward the shock (in the shock co-moving frame since the center of mass frame of the foreshock plasma differs from the shock frame) and may affect the onset and the saturation level of the Weibel instability. In addition, the current filaments producing the fields merge with time, so that $B$ and $\lambda$ change while being advected. These nonlinear feed-back effects are difficult to properly account for in a theoretical model; hence they are omitted in the current study. PIC simulations can help us to quantify the effects as well as to confirm or disprove our assumption that the shock and the foreshock do form a self-sustained, steady state structure.

\section{The self-similar foreshock}

The self-similar structure of the foreshock immediately follows from Eqs. (\ref{nCR}), (\ref{B-gamma}), (\ref{x-gamma}) and (\ref{gamma_r}). The magnetic field correlation length is proportional to the upstream distance from the shock,
\beq
\lambda(x)\sim x(2\xi_B),
\label{lambda-x}
\eeq
and its strength decreases with the distance as
\beq
B(x)\sim B_0 \left({x}/{x_0}\right)^{-\frac{s-1}{s+1}},
\label{B-x}
\eeq
where $B_0=\left(8\pi\xi_B m_pc^2n_0\gamma_0\right)^{1/2}$ and $x_0=\lambda_0/(2\xi_B)=\left(m_pc^2 \gamma_0/4\pi e^2 n_0\right)^{1/2}/(2\xi_B)$. In this estimate we neglected the advected fields $B(\gamma)$ as sub-dominant compared to $B(\gamma_r)$ for $\gamma>\gamma_r$. The $\epsilon_B$ parameter expresses the field energy normalized to the shock kinetic energy. The energy of cosmic rays is $U_{\rm CR}=\int n(\gamma/\gamma_0)(m_pc^2\gamma)\ d(\gamma/\gamma_0)\sim m_pc^2\gamma_0n_0$ and constitutes a fraction $\xi_{\rm CR}$ of the total shock energy, $U_{\rm sh}$. The efficiency of cosmic ray acceleration, $\xi_{\rm CR}$, can be as high as several tens percent, perhaps, up to $\xi_{\rm CR}\sim0.5$, as follows from the nonlinear shock modeling \citep{V+06,E+07}. The scaling of $\epsilon_B$ is:
\beq
\epsilon_B\sim\xi_{\rm CR}\xi_B\ \left({x}/{x_0}\right)^{-2\frac{s-1}{s+1}}.
\label{epsilonB-x}
\eeq

These scalings hold while the shock can be treated as planar and while the ISM magnetic fields are negligible compared to the Weibel-generated fields.  If the shock is relativistic, CR particles can occupy a narrow region in front of it. Assuming CR to propagate nearly at the speed of light, their front is ahead of she shock at the distance $\Delta r=c t_{\rm rel}=c(R/c-R/v_{\rm sh})\simeq R-R/[(1-1/2\Gamma^2_{\rm sh})]$ measured in the lab (observer) frame, that is at the distance $\sim\Delta r/\Gamma_{\rm sh} \sim R/(2\Gamma_{\rm sh})$ in the shock frame. Also, when the radial distance in the lab frame $\Delta r=x/\Gamma_{\rm sh}$ becomes comparable to the shock radius $\Delta r\sim R$ the curvature of the shock can no longer be neglected: the density of CR particles, which was assumed to be constant in our model, starts to fall as $\propto r^{-2}$. This leads to a steeper decline of $B$ with distance. Obviously, the first constraint is more stringent for a relativistic shock, whereas both are very similar (within a factor of two) for a non-relativistic shock. Hence we use the first constraint hereafter. Meanwhile, at some distance $X$, the Weibel-generated fields can become comparable to the ambient magnetic field, $B(X)\sim B_{\rm amb}$ and the Weibel instability ceases; here we used that the ambient field in the shock frame is $B_{\rm amb}\sim B_{\rm ISM,\perp}\Gamma_{\rm sh}\sim B_{\rm ISM}\Gamma_{\rm sh}$. PIC simulations \citep{Spit05} indicate that for low magnetizations $\sigma<0.01$, i.e., $B(X)/B_{\rm amb}>0.1$, the shock behaves as unmagnetized and the Weibel instability dominates. Although there is no a sharp threshold, one sees the Weibel instability to be suppressed for lower values of $B(X)/B_{\rm amb}$. To the order of magnitude, we set $B_{\rm ISM}\Gamma_{\rm sh}\sim B(X)\sim B_0(X/x_0)^{-(s-1)/(s+1)}$, therefore $X\sim x_0\left[B_0/(B_{\rm ISM}\Gamma_{\rm sh})\right]^{(s+1)/(s-1)}$. To conclude, the the scalings, Eqs. (\ref{lambda-x})--(\ref{epsilonB-x}), hold at $x\la x_{\rm max}$, where
\beq
x_{\rm max}={\rm Min}\left[R/(2\Gamma_{\rm sh}),\ X\right]={\rm Min}\left[R/(2\Gamma_{\rm sh}),\ x_0\left({B_0}/{B_{\rm ISM}\Gamma_{\rm sh}}\right)^{\frac{s+1}{s-1}}\right].
\label{xmax}
\eeq

The region filed with the magnetic field in front of the shock is large, so does the region where radiation emitted by the CR electrons. The power emitted by a relativistic electron in a magnetic field is $P_{B}=(4/3)\sigma_T c \gamma_e^2(B^2/8\pi)$, where $\sigma_T$ is the Thompson cross-section and $\gamma_e$ is the Lorentz factor of the emitting electron. This expression is accurate for both synchrotron and jitter radiation \citep{M00}. For the distribution of electrons (\ref{nCR}) homogeneously populating the foreshock, the power is dominated by the lowest energy particles with $\gamma_e\sim \epsilon_e(m_p/m_e)\gamma_p\sim \epsilon_e(m_p/m_e)\gamma_0\sim\epsilon_e(m_p/m_e)\Gamma_{\rm sh}$ (here $\epsilon_e$ is the efficiency of the electron heating) and the density $n_e\sim n_p\sim n_0\sim n_{\rm ISM}\Gamma_{\rm sh}$. Then $P_{\rm tot}=\int P_B(x)n_0dV$, where the volume element is $dV=4\pi R^2 dx$, so 
\beq
P_{\rm tot}\propto\int_0^{x_{\rm max}}(x/x_0)^{-2\frac{p-1}{p+1}}dx \sim x_0 \left(x_{\rm max}/x_0\right)^{\frac{3-s}{1+s}}.
\eeq 
Note that the co-moving radiated power {\em increases} with the foreshock thickness, $P_{\rm tot}\propto x_{\rm max}^{(3-s)/(1+s)}$, thus emission is not localized in a thin layer near the shock and is, in fact, dominated by large distances:
\beq
P_{\rm tot}\sim L_{\rm CR}\xi_B \gamma_e^2(2/3)\sigma_T n_0x_0 \left({x_{\rm max}}/{x_0}\right)^{\frac{3-s}{1+s}},
\label{Ptot}
\eeq
where $L_{\rm CR}=4\pi R^2(m_pc^2n_0\gamma_0)c$ is the kinetic luminosity of cosmic rays, which is a fraction $\xi_{\rm CR}<1$ of the total kinetic luminosity of a GRB, $L_{\rm CR}=\xi_{\rm CR} L_{\rm GRB}$, $R$ is the shock radius and the co-moving CR density is of order the density of the incoming ISM plasma, $n_0\sim n_{\rm ISM}\Gamma_{\rm sh}$. 
The foreshock electrons are radiating in the synchrotron regime: the jitter parameter $\delta$ \citep{M00,M+07}, which is the average deflection angle of an electron in the foreshock fields, $\theta_e\sim(eB(x)/\gamma_e m_ec)(\lambda(x)/c)$ over the radiation beaming angle, $\sim1/\gamma_e$, is always much larger than unity:
\beq
\delta(x)\sim eB(x)\lambda(x)/(m_ec^2)\sim (m_p/m_e)\gamma_0\sqrt{2\xi_B} (x/x_0)^{2/(s+1)} \gg1.
\eeq

Although consideration of the post-shock fields is beyond the scope of the present paper, we can estimate the magnetic field spectrum at and after the shock jump as long as dissipation is not playing a role. The magnetic field of different correlation scales created in the foreshock is advected toward the shock, so a broad spectrum is accumulated: 
\beq
B_\lambda\propto \lambda^{-\frac{s-1}{s+1}}\sim\lambda^{-0.091},
\label{Bspec}
\eeq
where Eqs. (\ref{lambda-x}) and (\ref{B-x}) were used and $s=p-1\sim1.2$ was assumed.

\section{The afterglow foreshock}

The relation between the shock radius $R$ and its Lorentz factor $\Gamma_{\rm sh}$ follows from a simple energy argument: the energy of an explosion is $E\sim(4\pi/3)R^3m_pc^2n_{\rm ISM}\Gamma_{\rm sh}^2$, therefore
\beq
R\sim(10^{18}\ {\rm cm})\ E_{52}^{1/3}n_{\rm ISM}^{-1/3}\Gamma_{\rm sh}^{-2/3}
\eeq
or $\Gamma_{\rm sh}\sim E_{52}^{1/2} n_{\rm ISM}^{-1/2} R_{18}^{-3/2}$, where $E_{52}=E/10^{52}\ {\rm erg}$ and similarly for other quantities.
The observed time of photons emitted at radius $R$ is the time since the very first photons (i.e, emitted at $R\sim0$) arrive, $t_{\rm obs}=R/c-R/v_{\rm sh}\simeq R/c-R/[c(1-1/2\Gamma^2_{\rm sh})]$, that is $t_{\rm obs}\sim R/(2\Gamma_{\rm sh}^2c)$. Using the equation for $R$, one gets
\begin{eqnarray}
\Gamma_{\rm sh}&\sim&3.7\ E_{52}^{1/8} n_{\rm ISM}^{-1/8} t_{\rm day}^{-3/8},\\
R&\sim&(4.2\times10^{17}\ {\rm cm})\ E_{52}^{1/4} n_{\rm ISM}^{-1/4} t_{\rm day}^{1/4}.
\label{Gamma-t}
\end{eqnarray}
Here we assumed a local GRB with $z=0$; to include the redshift time dilation is trivial.

The co-moving density is $n_0\sim n_{\rm ISM}\Gamma_{\rm sh}$ (assuming the CR efficiency $\xi_{\rm CR}\simeq0.5\sim1$) and the minimum Lorentz factor of CR protons is $\gamma_0\sim\Gamma_{\rm sh}$. Hence, the  length scale $x_0\sim\lambda_0/(2\xi_B)$ ($\lambda_0$ is the skin length) and the field $B_0$ in the shock co-moving frame become
\begin{eqnarray}
x_0&\sim& (2\times 10^7\ {\rm cm})\ n_{\rm ISM}^{-1/2}/(2\xi_B)\sim(10^9\ {\rm cm})\ n_{\rm ISM}^{-1/2}, \\
B_0&\sim& (0.2\ {\rm gauss})\ \xi_B^{1/2}n_{\rm ISM}^{1/2}\Gamma_{\rm sh}\sim (1~{\rm gauss})\ E_{52}^{1/2}R_{17}^{-3/2},
\label{x0B0}
\end{eqnarray}
where we assumed $\xi_B\sim0.01$. Assuming $p=2.2$ and the interstellar fields, $B_{\rm ISM}$, to be of order a microgauss, we estimate $X$ as 
\beq
X\sim x_0\left[ 0.2 (\xi_B n_{\rm ISM})^{1/2} B_{\rm ISM}^{-1}\right]^{\frac{s+1}{s-1}}
\sim 2\times10^{47}\ x_0\ n_{\rm ISM}^{5.5} B_{\rm ISM, -6}^{-11},
\eeq
independent of $\Gamma_{\rm sh}$. On the other hand, 
\beq
R/(2\Gamma_{\rm sh})\sim (5\times10^{17}\ {\rm cm})\ E_{52}^{1/3}n_{\rm ISM}^{-1/3}\Gamma_{\rm sh}^{-5/3}\ll X,
\eeq
indicating that the ambient field is relatively unimportant, even for very steep energy spectra $p\sim3.5$ rarely observed in prompt GRBs. The foreshock thickness is
\beq
x_{\rm max}\sim R/(2\Gamma_{\rm sh}) \sim 5\times10^8\ x_0\ E_{52}^{1/3}n_{\rm ISM}^{-1/3}\Gamma_{\rm sh}^{-5/3}.
\eeq
Therefore, the typical field in the foreshock is of sub-gauss strength: 
\beq
B(x_{\rm max})\sim(0.2\ {\rm gauss})\ E_{52}^{0.45}n_{\rm ISM}^{0.09}R_{18}^{-1.3}.
\eeq
This field is relatively large-scale, as it's co-moving correlation scale is
\beq
\lambda(x_{\rm max})\sim x_{\rm max}/(2\xi_B)\sim (5\times10^{17}\ {\rm cm})\ E_{52}^{-1/2} n_{\rm ISM}^{1/2} R_{18}^{5/2}.
\label{lambda-xmax}
\eeq
The power emitted by CR electrons from the foreshock amounts to 
\beq
P_{\rm tot}^{\rm obs}\sim (2\times10^{39}\ {\rm erg~s}^{-1})\ E_{52}^{1.6}L_{\rm CR, 45} n_{\rm ISM}^{-0.68}R_{18}^{-3.0}
\label{P-obs}
\eeq
in the observer's frame and is emitted at a peak (synchrotron) frequency
\beq
\nu_m^{\rm obs}\sim(10^{11}\ {\rm Hz})\ E_{52}^{2.0} n_{\rm ISM}^{-1.4} R_{18}^{-5.8},
\label{nu-obs}
\eeq
which corresponds to the IR band at about one day after the explosion, where $R(t)$ is given in Eq. (\ref{Gamma-t}), so that $\nu_m\propto t_{\rm day}^{-\frac{7s+17}{8(s+1)}}\propto t_{\rm day}^{-1.4}$.

\section{Discussion}

Here we presented a model of a self-similar foreshock produced by protons scattered by a relativistic shock into the unmagnetized or weakly magnetized external medium. It is immediately applicable to the external shock producing a GRB afterglow. The model predicts that a large region in front of the shock, of thickness of order the shock radius, shall be filled with relatively strong and large-scale magnetic fields. The upstream magnetic field strength and correlation length depend on the distance from the shock and the power-law index of accelerated protons (cosmic rays) and are given by Eqs. (\ref{lambda-x}), (\ref{B-x}), (\ref{x0B0}) in the shock co-moving frame. The overall energetics of the field is dominated by large distances; hence the average foreshock B-field is of sub-milligauss strength and is increasing toward the shock while its typical coherence length is of order of few percent of the shock radius and is decreasing toward the shock.

The result is interesting, especially in the light of observational constraints on the particle acceleration in GRB afterglows. It has been shown that a few-milligauss magnetic fields are needed in front of the shock in order to efficiently Fermi-accelerate the electrons to the energies required to produce the observed X-ray emission \citep{LW06}. Our model provides a possible and rather natural mechanism for generation of such fields in the far upstream medium. We also make a prediction that the shock-accelerated electrons will be radiating in the foreshock fields at a characteristic synchrotron frequency given by Eqs. (\ref{nu-obs}). For reference, $\nu_m\sim 10\ {\rm THz}$ at about a day after the burst. It is possible that nonlinear effects omitted in this analysis (see discussion below) may limit the field strength to lower values compared to the present analysis, so the synchrotron peak can move to cm/mm-wave band. We can speculate that the emission from the foreshock can form an emission region separate from the afterglow shock and show up in the X-ray/optical band in the early afterglow phase and in the radio at the very late times. 

The presented results can have interesting implications for the radiative efficiency of external shocks. Unlike internal shocks, where electron cooling is extremely fast and a thin shock layer of thickness of a hundred ion inertial lengths may be enough to produce the observed prompt emission \citep{MS09}, the Weibel shock model \citep{ML99} seems to face the efficiency problem when it is applied to an external shock. In such a shock, magnetic fields shall occupy a much larger region, perhaps the entire downstream region, in order for the shock to produce enough photons that will be observed as a delayed afterglow emission. This is not very likely (though not proved to be impossible, yet) provided that the small-scale fields generated {\em at} the shock can be subject to rapid dissipation. However, dissipation shall be of much lesser importance for the foreshock fields, which have a (much) larger coherence length, Eq. (\ref{lambda-x}). In the steady state, the fields generated in the upstream are advected to the shock and their strength is maintained against dissipation by the anisotropy of the continuously ``refreshed'' CR distribution. Hence, one can expect that the magnetic field near the shock will have the spectrum given in Eq. (\ref{Bspec}). Once these fields pass though the shock into the downstream, they are enhanced by the shock compression and begin to decay. Commonly, dissipation is proportional to the inverse gradient scale squared $\propto\partial_x^2\propto\lambda^{-2}$, so that the skin-length-scale fields may eventually disappear. However, the fields above a certain coherence length, $\lambda_{\rm diss}$, shall survive and fill up the post-shock medium. The mechanism of dissipation is not yet understood in detail, so it is premature to make any quantitative conclusions about $\lambda_{\rm diss}$, but it will certainly be much smaller than $\lambda(x_{\rm max})\lesssim R$, see Eq. (\ref{lambda-xmax}). Since $B_\lambda$ is a weak function of $\lambda$ one can speculate that relatively strong fields, perhaps of order tens or hundreds milligauss, to occupy the post-shock medium. 

 We want to note that a number of simplifying assumptions has been made in our analysis. In particular, {\em nonlinear feedback} effects of the upstream magnetic field on the particle distribution, on the shock structure and on Fermi acceleration were omitted. The inclusion of these effects is hardly possible in any analytical model. We also assumed that a steady state exists for the shock-foreshock system at hand. Apparently, it is not at all clear whether the steady state is at all possible or the system exhibits an intermittent behavior. One can envision a scenario in which the CRs overproduce upstream magnetic fields leading to enhanced particle scattering and the overall preheating of the ambient medium, which, in turn, can cause the shock to weaken,  disappear and then re-appear in a different place further upstream. Presently available 2D PIC simulations of an electron-position shock do show the upstream field amplification and no steady state has been achieved: both upstream and downstream fields continue to grow for the duration of the simulations \citep{KKSW08}. We argue that extensive PIC or/and hybrid simulations of a shock are imperative for further study.
 
Finally, we mention that our model compliments other models of the magnetic field generation. It is reasonable to expect that the field can be amplified by vortical motions produced by the Richtmeyer-Meshkov instability, if the ambient medium is clumpy or if the shock velocity is not perfectly uniform \citep{GM07,SG07,Milos+08}. On the other hand, if the ambient magnetic fields are strong enough, the fields can be generated via non-resonant \citep{BellLucek01,Bell04,PLM08} or resonant \citep{DM06,DM07,Z03,V+06,E+07} mechanisms, or both \citep{Kirk+08}.

\ack

The authors thank colleagues at IKI and RRC ``KI" for discussions. This work has been supported by NSF grant AST-0708213, NASA ATFP grant NNX-08AL39G, Swift Guest Investigator grant NNX-07AJ50G and DOE grant  DE-FG02-07ER54940.

\begin{figure}
\includegraphics[width=5.5in]{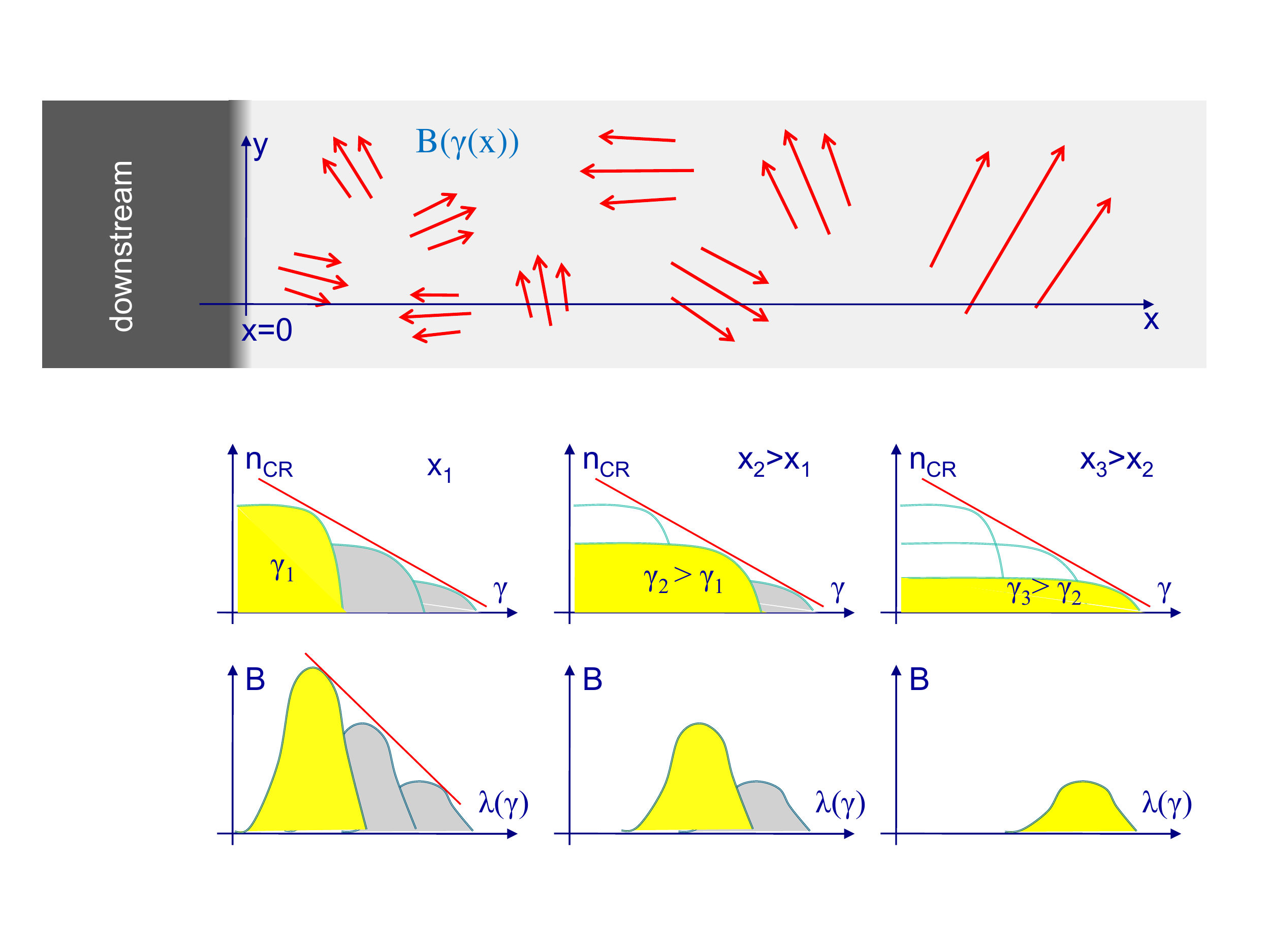}
\caption{A schematic representation of the foreshock magnetic fields: the coherence length is increasing with the upstream distance. Below are schematic graphs showing variation of the spectrum of the streaming part of cosmic rays and the corresponding self-generated fields (highlighted).
\label{f1}}
\end{figure}

\end{document}